\documentclass[letterpaper, 10 pt, conference]{ieeeconf}  % Comment this line out if you need a4paper
\IEEEoverridecommandlockouts
\usepackage{graphics} % for pdf, bitmapped graphics files
\usepackage{epsfig} % for postscript graphics files
\usepackage{amsfonts}

\usepackage{times} % assumes new font selection scheme installed
\usepackage[utf8]{inputenc}
\usepackage[english]{babel}
\usepackage{amsmath}
\usepackage{amssymb}
\usepackage{hyperref}
\usepackage{cleveref}
\usepackage{rotating}
\usepackage{cite}
\usepackage{graphicx}
\newcommand{\ols}[1]{\mskip.5\thinmuskip\overline{\mskip-.5\thinmuskip {#1} \mskip-.5\thinmuskip}\mskip.5\thinmuskip} % overline short
\newcommand{\olsi}[1]{\,\overline{\!{#1}}} % overline short italic
\makeatletter
\newcommand\closure[1]{
  \tctestifnum{\count@stringtoks{#1}>1} %checks if number of chars in arg > 1 (including '\')
  {\ols{#1}} %if arg is longer than just one char, e.g. \mathbf{Q}, \mathbf{F},...
  {\olsi{#1}} %if arg is just one char, e.g. K, L,...
}
\long\def\count@stringtoks#1{\tc@earg\count@toks{\string#1}}
\long\def\count@toks#1{\the\numexpr-1\count@@toks#1.\tc@endcnt}
\long\def\count@@toks#1#2\tc@endcnt{+1\tc@ifempty{#2}{\relax}{\count@@toks#2\tc@endcnt}}
\def\tc@ifempty#1{\tc@testxifx{\expandafter\relax\detokenize{#1}\relax}}
\long\def\tc@earg#1#2{\expandafter#1\expandafter{#2}}
\long\def\tctestifnum#1{\tctestifcon{\ifnum#1\relax}}
\long\def\tctestifcon#1{#1\expandafter\tc@exfirst\else\expandafter\tc@exsecond\fi}
\long\def\tc@testxifx{\tc@earg\tctestifx}
\long\def\tctestifx#1{\tctestifcon{\ifx#1}}
\long\def\tc@exfirst#1#2{#1}
\long\def\tc@exsecond#1#2{#2}
\makeatother

\title{\LARGE \bf
MRAC with Memory for Switched Linear Systems }
\author{Pritesh Patel$^{1}$, Sayan Basu Roy$^{2}$ and Shubhendu Bhasin $^{3}$% <-this % stops a space
\thanks{$^{1}$Pritesh Patel is a PhD student in Control and Automation, Electrical Engineering, Indian Institute of Technology Delhi, India
        {\tt\small priteshpatel.iitd@gmail.com}}%
\thanks{$^{2}$Sayan Basu Roy is Assistant Professor at  Indraprastha Institute of Information Technology Delhi, India
        {\tt\small sayan@iiitd.ac.in}}%
\thanks{$^{3}$Shubhendu Bhasin is Professor at  Indian Institute of Technology Delhi, India
        {\tt\small sbhasin@ee.iitd.ac.in}}}

\begin{document}
\maketitle
\thispagestyle{empty}
\pagestyle{empty}

%%%%%%%%%%%%%%%%%%%%%%%%%%%%%%%%%%%%%%%%%%%%%%%%%%%%%%%%%%%%%%%%%%%%%%%%%%%%%%%%
\begin{abstract}
This work proposes a switched model reference adaptive control (S-MRAC) architecture for a multi-input multi-output (MIMO) switched linear system with memory for enhanced learning. A salient feature of the proposed method that separates it from most previous results is the use of memory that store the estimator states at switching and facilitate parameter learning during both active and inactive phases of a subsystem, thereby improving the tracking performance of the overall switched system. Specifically, the learning experience from the previous active duration of a subsystem is retained in the memory and reused when the subsystem is inactive and when the subsystem becomes active again. Parameter convergence is shown based on an intermittent initial excitation (IIE), which is significantly relaxed than the classical persistence of excitation (PE) condition. A common Lyapunov function is considered to ensure closed-loop stability with S-MRAC. Further under IIE, the exponential stability of tracking 
and parameter estimation error dynamics are guaranteed.  
\end{abstract}.    
\section{INTRODUCTION}

 Hybrid systems consist of continuous and discrete dynamics such that continuous dynamics take values from discrete states as input and vice versa. Many dynamical systems in practice are hybrid, e.g., the motion of the automobile, power grids, etc. Continuous-time systems with discrete switching events are referred to as switched systems, a subclass of hybrid systems that have a wide variety of applications ranging from aerospace, robotics, and transportation \cite{liberzon2003switching},\cite{yuan2016adaptive}.

Control of switched systems under parametric uncertainty is a challenging problem and has been an active area of research. Model reference adaptive control (MRAC) is often used to handle parametric variations, and uncertainties \cite{landau1979adaptive}. Simplicity in design and implementation of MRAC has motivated the extension to switched systems \cite{6263274, di2008novel, sang2011adaptive, bernardo2013model, 7782779, 9642040,9521772, kersting2017direct,5991336,wu2015adaptive}. In particular, in \cite{7782779}, adaptive tracking of the switched system state to the reference system state is proved by using a time-varying gain matrix in multiple quadratic Lyapunov functions. In\cite{6263274} and \cite{di2008novel}, minimum control synthesis (MCS) is used for asymptotic tracking of the multimodal piecewise affine system to the states of a multimodal piecewise affine
reference model. Extension of the algorithm to the discrete-time system is described in \cite{bernardo2013model}. In \cite{kersting2017direct}, direct and indirect MRAC strategies are proposed with and without using a common Lyapunov function. In \cite{9642040}, a novel indirect MRAC is developed using a barrier Lyapunov function and with multiple Lyapunov functions where an average dwell time constraint is required to establish the boundedness of all closed-loop signals and PE condition is required to ensure parameter convergence. PE condition is restrictive because signal is required to have
sufficient energy for the entire time-span \cite{Roy2018CombinedConvergence}. In \cite{wu2015adaptive}, state tracking MRAC for the switched system with parametric uncertainties is analyzed where switching between subsystem and controller is asynchronous. A detailed study of direct and indirect MRAC for the piecewise affine (PWA) systems is presented in \cite{7890417} where stability is analyzed using common quadratic Lyapunov function (CQLF) and multiple quadratic Lyapunov functions (MQLF), and parameter convergence is shown using the PE condition. The dwell time expression is derived using the fact that the parameter learning stops in the inactive phase.
In \cite{sang2011adaptive}, adaptive control of the piecewise linear system is presented for state tracking case where tracking error is shown to be bounded for all time and parameter convergence is shown under the PE condition.

In all the approaches mentioned above, multiple reference models are considered, where synchronization of the switching signal for switched system and reference system is a critical concern. Adaptive control of switched systems with asynchronous switching is studied in \cite{7959189} and \cite{xie2017model}, where a single reference model is considered to simplify the analysis, and hence the requirement of the synchronous switching becomes irrelevant. One advantage of using a single reference model is that the dwell time requirement using CQLF and MQLF is the same, i.e., in both cases the switched system is stable for arbitrary fast switching \cite[Corollary 1]{sang2011adaptive}. A common feature in switched system literature is that the parameter learning stops during the inactive period of the subsystem, other features include the requirement of a large amount of data recording while the subsystem is active \cite{kersting2014concurrent}, and the requirement of the persistence of excitation (PE) condition on regressor for parameter convergence.

The proposed work presents a switched MRAC architecture for the switched linear systems, where information about the unknown subsystem parameter, collected during the active phase of the subsystem, is stored in the memory and then used for parameter learning in the inactive period and later when the same subsystem is active again. The stability of the overall switched system is analyzed using a CQLF. No dwell time constraint is required for the switching signal. Online parameter estimation of the switched affine system is studied in the authors' previous work \cite{https://doi.org/10.48550/arxiv.2204.03338} where parameter learning in the inactive period of the subsystem is shown. A similar approach for the switched MRAC is followed in this work to enable parameter learning in the inactive phase. Parameter convergence of the switched system parameters is shown if the regressor satisfies the intermittent initial excitation (IIE), which is a more generalized version of initial excitation (IE) proposed in \cite{roy2017uges, roy2017combined} and less stringent than the persistence of excitation (PE) condition mentioned in the literature for parameter convergence. Major contributions of this work are:
\begin{itemize}
    \item Parameter learning in the inactive period of the subsystem (unlike \cite{kersting2017direct, sang2011adaptive, yuan2016adaptive}), which improves the overall tracking performance of the switched system.
    \item Parameter convergence without the requirement of PE condition (unlike \cite{kersting2017direct, sang2011adaptive, yuan2016adaptive, 9642040}).
\end{itemize}
\textbf{\textit{Notations}}: $||\bullet||$ denotes the Euclidean norm of a vector; $tr\{\bullet\}$ is the trace of a matrix; $I_n$ denotes the identity matrix of order $n$; $\otimes$ denotes the matrix Kronecker product; $vec(Z)\in\mathbf{R}^{ab}$ denotes the vectorization of a matrix $Z\in\mathbf{R}^{a\times b}$ obtained by stacking the columns of the matrix $Z$.
\section{PROBLEM FORMULATION AND PRELIMINARIES}\label{s2}
\subsection{System description}
 Consider the following uncertain switched LTI system
 \begin{equation}\label{eq1}
     \dot{x}(t)=A_{\sigma(t)}x(t)+B_{\sigma(t)}u(t), x(t_0)=x_0,\hspace{0.2cm} 
 \end{equation}
 where $x\in \mathbf{R}^n$ is the state vector, $u\in \mathbf{R}^m$ is the control input, $t\in[t_0,\infty),\hspace{0.2cm} t_0\geq 0 $, $\sigma : [0, \infty) \rightarrow \mathbf{S}$ denotes a piecewise constant switching signal, where $\mathbf{S}=\{1, 2, 3,...,M\}$,  $A_{i}\in\mathbf{R}^{n\times n}$ denotes the system matrix which is unknown and $B_{i} \in \mathbf{R}^{n \times m}$ is known input matrix for the $i^{th}$ subsystem ($i\in \mathbf{S}$). The system starts from an initial time $t_0$, and let $t_k\hspace{0.1cm}(k=1,2,3,...)$ denote the time instants when the system switches from one subsystem to another based on the switching signal $\sigma(t)$, which is discontinuous at the switching instants and has a constant value between two consecutive switching instants. 
  
\textbf{\textit{Assumption 1}}: Control matrix $B_i, \forall i \in \mathbf{S}$ has full column rank so that $(B_i^TB_i)^{-1}$ exists $\forall i \in \mathbf{S}$.  
   
For all subsystems of the switched system (\ref{eq1}), a common reference representing the desired closed-loop behavior is chosen as
  \begin{equation}\label{ref_model}
      \dot{x}_m(t)=A_{m}x_m(t)+B_{m}r(t)
  \end{equation}
where $x_m\in \mathbf{R}^n$ is reference model state, and $r\in \mathbf{R}^m$ denotes a bounded piecewise constant reference input signal. $A_{m} \in \mathbf{R}^{n \times n}$ and $B_{m}\in \mathbf{R}^{n \times m}$ are known, and $A_{m}$ is Hurwitz.
\subsection{Adaptive Control Law}
A certainty equivalence state-feedback adaptive control law is developed as
\begin{equation}\label{control_input}
    u(t)=\hat{K}_{xi}^Tx(t)+K_{ri}^Tr(t), \hspace{0.5cm} \forall i \in \mathbf{S}
\end{equation}
where $ \hat{K}_{xi}\in \mathbf{R}^{n \times m}$ is the time varying control gain, which is also denoted as direct estimates of the true control parameter and $K_{ri}\in \mathbf{R}^{m \times m}$ is constant control gain.  Substituting (\ref{control_input}) in (\ref{eq1}) yields
\begin{equation}\label{model1}
\dot{x}=(A_i+B_{i}\hat{K}_{xi}^T)x(t)+B_{i}K_{ri}^Tr(t), \hspace{0.5cm} \forall i \in \mathbf{S}
\end{equation}
To facilitate the design objective of making system (\ref{model1}) track the reference model (\ref{ref_model}), the following matching condition is introduced\\
\textit{\textbf{Assumption 2}}: There exists constant matrices $K_{xi}\in \mathbf{R}^{n \times m}$ and $K_{ri} \in \mathbf{R}^{m \times m}$,  $\forall i \in \mathbf{S}$, such that
\begin{subequations}
\begin{align}
    A_i+B_iK_{xi}^T&=A_{m}\label{5a}\\
    B_iK_{ri}^T&=B_{m}\label{5b}
\end{align}
\end{subequations}
Using (\ref{5a}) and (\ref{5b}), closed-loop system in (\ref{model1}) can be written as
\begin{equation}\label{eq_6}
    \dot{x}=A_{m}x_m+B_{m}r+B_{i}\tilde{K}^T_{xi}x+B_{i}K^T_{ri}r, \hspace{0.5cm} \forall i \in \mathbf{S}
\end{equation}
where $\tilde{K}^T_{xi}\triangleq \hat{K}_{xi}^T-K_{xi}^T$. The tracking error is defined as
\begin{equation} \label{tracking_error}
    e(t)=x(t)-x_m(t)
\end{equation}
Using (\ref{ref_model}), (\ref{eq_6}) and (\ref{tracking_error}), the error dynamics is obtained as
\begin{equation}\label{eq_8}
    \dot{e}=A_{m}e+B_{i}\tilde{K}^T_{xi}x, \hspace{0.5cm} \forall i \in \mathbf{S}
\end{equation}
The control law in (\ref{control_input}) can be written as
\begin{equation}\label{eq_42}
    u(t)=u_k(t)+u_e(t)
\end{equation}
where $u_k=K_{ri}^Tr$  
and $u_e=Z(x)\hat{\phi}_{i}(t)$
with $Z\in \mathbf{R}^{m\times mn}$ is specified as $Z=I_m\otimes x^T$
and $\hat{\phi}_{i} \in \mathbf{R}^{mn}$ is the vector of direct estimate of the control parameter, consisting of $\hat{K}_{xi}$, defined as $\hat{\phi}_{i}\triangleq 
    vec(\hat{K}_{xi}), \forall i \in \mathbf{S}.$ Using (\ref{eq_42}), the error dynamics in (\ref{eq_8}) can be written as 
\begin{equation}\label{eq45}
    \dot{e}=A_me+B_{i}Z\tilde{\phi}_{i}, \hspace{0.5cm} \forall i \in \mathbf{S}
\end{equation}
where $\tilde{\phi}_{i}\triangleq \hat{\phi}_{i}-\phi_{i}$, and the true controller parameter vector is defined as
\begin{equation}
    \phi_{i}\triangleq 
    vec({K_{xi}}), \hspace{0.5cm} \forall i \in \mathbf{S}
\end{equation}
\subsection{Control Objective}
The objective is to design a control law $u(t)$ and parameter estimation law for $\dot{\hat{\phi}}_i(t)$  such that $x(t)\rightarrow x_m(t)$ and $\hat{\phi}_i(t)\rightarrow \phi_i$ as $t\rightarrow \infty, \forall i \in \mathbf{S}$. 
\subsection{Preliminary Definitions}
Consider the following definition for a general signal and system  pair ($\phi(t,x),f(t,x)$), which is subsequently used in Assumption 2.

\textit{\textbf{Definition}}: A function $\varphi(t,x)\in \mathbf{R}^{p\times q}$, where $p>q>0$, is called uniformly \textit{intermittent} IE (u-IIE) w.r.t. $\dot{x}=f(t,x)$ and indicator function $\mathfrak{I}(t) :[0,\infty)\rightarrow \{0,1\}$ if  $\exists \hspace{0.1cm}\alpha, T>0$ such that $\forall (t_0,x_0)\in \mathbf{R}_{\geq 0}\times \mathbf{R}^n $, all corresponding solutions satisfy
\begin{equation}
    \int_{t_{0}}^{t_{0}+T}\mathfrak{I}(\tau)\varphi(\tau, x(\tau,t_0,x_0))^T\varphi(\tau,x(\tau,t_0,x_0))d\tau\geq \alpha I_q 
\end{equation}
where $\mathfrak{I}(t)\in \{0,1\}$ is an indicator function which has a value of 1 when the system is active and 0 when the system is inactive.
\subsection{First Layer Filters}
Consider the following filter equations
\begin{subequations}
\begin{align}
\dot{e}_{df}&=-k_fe_{df}+\dot{e},\hspace{0.6cm}&e_{df}(t_k)=\mathbf{S_{e_{{df}_{\sigma}}}}_{(t_k)}\label{eq13a} \\
\dot{e}_{f}&=-k_fe_{f}+e,\hspace{0.6cm}&e_{f}(t_k)=\mathbf{S_{e_{f_{\sigma}}}}_{(t_k)}\label{eq13b} \\
\dot{u}_{ef}&=-k_fu_{ef}+u_e, \hspace{0.6cm}&u_{ef}(t_k)=\mathbf{S_{u_{{ef}_{\sigma}}}}_{(t_k)} \label{eq13c} \\
\dot{Z_{f}}&=-k_fZ_{f}+Z,\hspace{0.6cm}&Z_{f}(t_k)=\mathbf{S_{Z_{f_{\sigma(t_k)}}}}_{(t_k)} \label{eq13d}
\end{align}
\end{subequations}
$\hspace{4cm}k=0,1,2,3,... \hspace{0.1cm} \& \hspace{0.1cm} \sigma(t)\in \mathbf{S}$\\
where $e_{df}(t)\in \mathbf{R}^{n}$, $e_f(t)\in \mathbf{R}^n, u_{ef}(t)\in \mathbf{R}^{m}, Z_f(t) \in \mathbf{R}^{m \times mn}$ denote the filtered error derivative, filtered error, filtered control input and filtered regressor respectively, and $k_f>0$ is a scalar gain introduced to stabilize the filters.
Further, $\mathbf{S_{e_{df}}}_i \in \mathbf{R}^{n}$, $\mathbf{S_{e_f}}_i \in \mathbf{R}^{n}$, $\mathbf{S_{u_{ef}}}_i \in \mathbf{R}^{m}$ and $\mathbf{S_{Z_f}}_i \in \mathbf{R}^{m \times mn}$
 denote the $i^{th}$ element of the memory stacks $\mathbf{S_{e_{df}}}$, $ \mathbf{S_{e_f}}$, $ \mathbf{S_{u_{ef}}}$ and $ \mathbf{S_{Z_f}} $ respectively, which store the filter value at the switch-out instants corresponding to that $i^{th}$ subsystem ($i \in \mathbf{S})$. The memory stacks are defined as $\mathbf{S_{e_{df}}}\triangleq[\mathbf{S_{e_{{df}_1}}},\mathbf{S_{e_{{df}_2}}},...\mathbf{S_{e_{{df}_M}}}]$, $\mathbf{S_{e_f}}\triangleq [\mathbf{S_{e_{f_1}}},\mathbf{S_{e_{f_2}}},...\mathbf{S_{e_{f_M}}}]$, $\mathbf{S_{u_{ef}}}\triangleq [\mathbf{S_{u_{{ef}_1}}},\mathbf{S_{u_{{ef}_2}}},...\mathbf{S_{u_{{ef}_M}}}]$, $\mathbf{S_{Z_f}}\triangleq [\mathbf{S_{Z_{f_1}}},\mathbf{S_{Z_{f_2}}},...\mathbf{S_{Z_{f_M}}}]$,  and are initialized to zero, i.e. $\mathbf{S_{e_{{df}_i}}}(t_0)=0,\hspace{0.1cm}\mathbf{S_{e_{f_i}}}(t_0)=0, \hspace{0.1cm}\mathbf{S_{u_{{ef}_i}}}(t_0)=0, \hspace{0.1cm}\mathbf{S_{Z_{f_i}}}(t_0)=0\hspace{0.1cm}, i\in \mathbf{S}$. The memory stacks are populated using the following logic
\begin{subequations}
\begin{align}
    \mathbf{S_{e_{df}\sigma}}_{(t_k^-)}&={e_{df}}(t^-_k), \hspace{0.5cm} k=1,2,3,...\label{eq14a}\\
\mathbf{S_{e_f\sigma}}_{(t_k^-)}&={e_f}(t^-_k), \hspace{0.5cm} k=1,2,3,...\label{eq14b}\\
\mathbf{S_{u_{ef}\sigma}}_{(t_k^-)}&={u_{ef}}(t^-_k), \hspace{0.5cm} k=1,2,3,...\label{eq14c}\\
\mathbf{S_{Z_f\sigma}}_{(t_k^-)}&={Z_f}(t^-_k), \hspace{0.5cm} k=1,2,3,...\label{eq14d}
\end{align}
\end{subequations}
where $t_k\hspace{0.1cm}(k=1,2,3...)$ denotes the switching instant \& $t^-_k$ denotes the time just before the switching instant $t_k$.\\
\textit{\textbf{Remark}}: The equations (\ref{eq13a})-(\ref{eq13d}) represent the continuous filter dynamics and the discrete reset of filter states at switching instants $t_k\hspace{0.1cm}(\forall k \in \mathbf{N})$. At each switching instant $t_k$, when the system switches from, say, subsystem $q$ to subsystem $r\hspace{0.1cm} (q,r\in\mathbf{S}),$ the filter states $e_{df}(t_k^-),e_f(t_k^-),u_{ef}(t_k^-)$ and $Z_f(t_k^-)$ corresponding to the unknown parameter of subsystem $q$ are recorded in the memory stack at locations $\mathbf{S_{e_{df}}}_q$, $\mathbf{S_{e_f}}_q,\mathbf{S_{u_{ef}}}_q $ and $\mathbf{S_{Z_f}}_q$ respectively. These stored filter values are used when subsystem $q$ is switched OFF and later recalled whenever subsystem $q$ is switched back ON.\\
Filter equation (\ref{eq13a}) can be solved explicitly as
\begin{equation}\label{eq8}
    e_{df}(t)=e_{df}(t_k)+e^{-k_ft}\int_{t_{k}}^{t}e^{k_f\tau}\dot{e}(\tau)d\tau,\hspace{0.1cm}
    e_{df}(t_k)=\mathbf{S_{e_{df}\sigma}}_{(t_k)}
\end{equation}
    \hspace{4cm}$t\in [t_k,t_{k+1}), k=0,1,2,3,...$
    Using the by parts rule of integration, (\ref{eq8}) can be written as
\begin{multline}\label{eq9}
    e_{df}(t)=e_{df}(t_k)+e(t)-\exp\{-k_f(t-t_k)\}e(t_k)\\-k_f(e_f(t)-e_f(t_k)), \hspace{0.1cm} t\in [t_k,t_{k+1}), k=0,1,2,3,...
\end{multline}
Since $e(t), e_{f}(t)$ are known quantities, $e_{df}(t)$ can be obtained online using (\ref{eq9}), instead of the unimplementable form in (\ref{eq13a}).

Further, explicitly solving (\ref{eq13b})-(\ref{eq13d}) and using (\ref{eq45}) in (\ref{eq8}), the following relation is deduced\footnote{Assumption 1 is used in (\ref{eq17c}).}
\begin{subequations}
\begin{align}
e_{df}(t)&=A_me_f(t)+B_i(u_{ef}(t)-Z_f(t)\phi_i)
\intertext { Subsequent manipulations yield }
   \underbrace{e_{df}-A_{me_f}}_{h}&=B_i(u_{ef}-Z_{f\phi_i})\\
    u_{ef}-Z_{f\phi_i}&=\underbrace{(B_i^TB_i)^{-1}B_i^Th}_{h_{B_i}} \label{eq17c}\\
    Z_f\phi_i&=\underbrace{u_{ef}-h_{B_i}}_{u_{ei}}\\
Z_f(t)\phi_i&=u_{ei}(t), \hspace{0.5cm} \forall i\in \mathbf{S}, t\geq t_0 \label{eq22b}
\end{align}
\end{subequations}

\subsection{Second Layer Filters}
To obviate the need for PE, another layer of filter equations, inspired from \cite{kreisselmeier1977adaptive}, are used
\begin{subequations}
\begin{align}
\dot{Q}&=-k_sQ+Z_f^TZ_f, \hspace{0.6cm} Q(t_k)=\mathbf{S_{Q_{\sigma}}}_{(t_k)} \label{eq11}\\
\dot{G}&=-k_sG+Z_f^Tu_{ei}, \hspace{0.6cm} G(t_k)=\mathbf{S_{G_{\sigma}}}_{(t_k)} 
\end{align}
\end{subequations}
$$\hspace{0.5cm}k=0,1,2,3,... \hspace{0.1cm} \& \hspace{0.1cm} \sigma(t)\in \mathbf{S}$$
where $k_s>0$ is a scalar gain and $Q(t)\in \mathbf{R}^{mn\times mn}$ denotes the double-filtered regressor and $G(t)\in \mathbf{R}^{mn}$, $\mathbf{S_{Q_{i}}} \in \mathbf{R}^{mn \times mn}$, $\mathbf{S_{G_{i}}} \in \mathbf{R}^{mn}$
 denote the $i^{th}$ element of memory stack $\mathbf{S_{Q}}$ and $ \mathbf{S_{G}}$, which store the filter value at the switch out instants corresponding to that $i^{th}$ subsystem ($i \in \mathbf{S})$. The resulting memory stacks are defined as $\mathbf{S_{Q}}=[\mathbf{S_{Q_1}},\mathbf{S_{Q_2}},...\mathbf{S_{Q_M}}]$, $\mathbf{S_{G}}=[\mathbf{S_{G_1}},\mathbf{S_{G_2}},...\mathbf{S_{G_M}}]$. The memory stacks are populated using the following logic
\begin{subequations}
\begin{align}
    \mathbf{S_{Q\sigma}}_{(t_k^-)}&={Q}(t^-_k), \hspace{0.5cm} k=1,2,3,...\\
    \mathbf{S_{G\sigma}}_{(t_k^-)}&={G}(t^-_k), \hspace{0.5cm} k=1,2,3,...
    \end{align}
\end{subequations}
The following relation can be deduced with the help of (\ref{eq22b}) as
\begin{equation}
    G(t)=Q(t)\phi_{i}, \hspace{0.8cm} \forall t\geq t_0\end{equation}
From (\ref{eq11}), the square matrix $Q(t)$ can be expressed as
\begin{equation}\label{eq28}
    Q(t)=\underbrace{Q(t_k)}_{\geq 0}+\underbrace{e^{-k_st}}_{\geq 0}\int_{t_{k}}^{t}\underbrace{e^{k_sr}}_{\geq1}\underbrace{Z_f(r)^TZ_f(r)}_{\geq 0}dr,\hspace{0,2cm}
\end{equation}
\hspace{1cm}$Q(t_k)=\mathbf{S_{Q_{\sigma}}}_{(t_k)},t\in [t_k,t_{k+1}),\hspace{0.2cm} k=0,1,2,3,...$\\
Using (\ref{eq28}), the following property can be derived.\\
\textit{\textbf{Property 1}}. $Q(t)$ is a positive semi-definite function of time i.e. $Q(t)\geq 0, \hspace{0.2cm} \forall t\geq t_0.$

\subsection{Parameter Estimation Design}\footnote{$C_e$ is the extra term in parameter estimation law compared to the previous work in \cite{https://doi.org/10.48550/arxiv.2204.03338}.}
The switched parameter estimation law for subsystem $i\in \mathbf{S}$ is proposed as
\begin{equation}\label{eq_para_est}
    \dot{\hat{\phi}}_i=\begin{cases}\Gamma_{\phi_i}\underbrace{(C_e+C_{li}+C_{lli}+{{s_{i}}}{C_{sw i})}}_{C_{\tilde{\phi}_{i}}},\hspace{0.8cm} \text{when} \hspace{0.2cm} \sigma(t)=i\\
    
    \Gamma_{\phi_i}\underbrace{(\closure{C}_{li}+\closure{C}_{lli}+{s_{i}}C_{sw i})}_{\closure{C}_{\tilde{\phi}_{i}}},\hspace{1.5cm} \text{when} \hspace{0.2cm} \sigma(t)\neq i
    \end{cases}
\end{equation}
\begin{equation*}
    \hspace{4 cm}\hat{\phi}_i(t_0)=\hat{\phi}_{i0},\hspace{0.1cm} \forall i \in \mathbf{S},\hspace{0.1cm} t\geq t_0
\end{equation*}
where $\Gamma_{\theta i} \in \mathbf{R}^{mn\times mn}$ is a positive-definite learning gain matrix. The terms in (\ref{eq_para_est}) are given by
\begin{subequations}
\begin{align}
    C_e & \triangleq Z^TB_i^TP^Te\\
    C_{li} &\triangleq k_{li} Z_f^T(t)[u_{ei}(t)-Z_f(t)\hat{\phi}_i(t)] \label{eq21}\\
    C_{lli}&\triangleq k_{lli}[G(t)-Q(t)\hat{\phi}_{i}(t)] \label{eq22}\\
    C_{swi}&\triangleq k_{swi}[\mathbf{S_{\closure{G}_i}}-\mathbf{S_{\closure{Q}_i}}\hat{\phi}_{i}(t)]\label{eq_s} \\
    \closure{C}_{li}&\triangleq k_{li} \mathbf{S^T_{Z_{f_i}}}[\mathbf{S_{e_{{df}_i}}}-\mathbf{S_{Z_{f_i}}}\hat{\phi}_{i}(t)]\label{eq21p}\\
    \closure{C}_{lli}&\triangleq k_{lli}[\mathbf{S_{G_i}}-\mathbf{S_{Q_i}}\hat{\phi}_{i}(t)]\label{eq22p}
\end{align}
\end{subequations}
\hspace{6cm}$\forall t \geq t_0,\forall i\in \mathbf{S}$\\
The piecewise-constant switching signal ${s_{i}}(t)\in \mathbf{R}$ is defined as
\begin{equation}\label{eq26}
    {s_{i}}(t)=\begin{cases}
    0 &\hspace{0.5cm} \text{for}\hspace{0.5cm} t\in [t_{0},\hspace{0.1cm}t_{0}+T_{i})\\
    1& \hspace{0.5cm} \text{else}
    \end{cases}
\end{equation}
\hspace{6cm} $\forall i \in \mathbf{S}$\\
where $k_{li}, k_{lli}, k_{swi}>0$ are scalar gains and $\mathbf{S_{\closure{G}_i}}\in \mathbf{R}^{mn},\hspace{0.1cm}\mathbf{S_{\closure{Q}_i}} \in \mathbf{R}^{mn\times mn}$ and $\mathbf{S_{s_i}} \in \mathbf{R}$ denote the $i^{th}$ element of memory stack $\mathbf{S_{\closure{G}}},\hspace{0.1cm}\mathbf{S_{\closure{Q}}}$ and $\mathbf{S_s}$ corresponding to that $i^{th}$ subsystem $(i\in \mathbf{S}$). The resulting memory stacks are defined as $\mathbf{S_{\closure{G}}}=[\mathbf{S_{\closure{G}_1}},\mathbf{S_{\closure{G}_2}},...\mathbf{S_{\closure{G}_M}}],\hspace{0.1cm} \mathbf{S_{\closure{Q}}}=[\mathbf{S_{\closure{Q}_1}},\mathbf{S_{\closure{Q}_2}},...\mathbf{S_{\closure{Q}_M}}]$ and $\mathbf{S_s}=[\mathbf{S_{s_1}},\mathbf{S_{s_2}},...\mathbf{S_{s_M}}]$, $\mathbf{S_{s_i}}=1$ indicates IIE condition for subsystem $i \in \mathbf{S}$ is achieved. The memory stacks are populated using the following logic
\begin{subequations}
\begin{align}
\mathbf{S_{\closure{G}_i}} & \triangleq G_{i}(t_{0}+T_{i})\\
\mathbf{S_{\closure{Q}_i}} & \triangleq Q_{i}(t_{0}+T_{i}) \label{eq_Q_bar}\\
\mathbf{S_{s_i}} &  \triangleq s_{i}(t)\hspace{1.5 cm} i\in \mathbf{S} \label{eq_S_s}
\end{align}
\end{subequations}
% \textbf{\textit{Remark}}: In \cite{https://doi.org/10.48550/arxiv.2204.03338}, only the subsystem parameter identification is studied whereas in this paper design of control input along with the subsystem parameter identification for switched MRAC is analyzed.

\textit{\textbf{Assumption 3}} : The regressor $Z_f(t,x)$ is u-IIE w.r.t. indicator functions $\mathfrak{I}_i(t),\ \forall i\in\mathbf{S}$ and the tracking and parameter estimation error dynamics in (\ref{eq45}) and (\ref{eq_para_est}) respectively (as per Definition) with degree of excitation $\gamma_i$ and  i.e.  
\begin{multline}\label{eq15}
    \int_{t_{0}}^{t_{0}+T_i}\mathfrak{I}_i(\tau)Z_f(\tau, x(\tau))^TZ_f(\tau,x(\tau))d\tau\geq \gamma_i I_{mn}, \\\hspace{1cm} i\in \mathbf{S}
\end{multline}
where $\mathfrak{I}_i(t)$ is an indicator function for the $i^{th}$ subsystem defined as
\begin{equation}
    \mathfrak{I}_i(t)=\begin{cases}
    1, & \text{when} \hspace{0.2cm}\sigma(t)=i\\
    0, & \text{when} \hspace{0.2cm} \sigma(t)\neq i
    \end{cases}
\end{equation}
\textit{\textbf{Lemma 1}: A necessary and sufficient condition for the regressor $Z_f(t,x)$ to be IIE for subsystem $i$ is that $Q(t_{0}+T_i)$ is a positive definite (PD) matrix, where $t_0+T_i \in \bar{t}_i$ and $\bar{t}_i=\{t\hspace{0.1cm} |\hspace{0.1cm} \sigma(t)=i\}$.}\\
\textit{\textbf{Lemma 2}\footnote{Proofs of Lemmas 1-2 are available in \cite{https://doi.org/10.48550/arxiv.2204.03338}.}: If $Q(t_{0}+T_i)$ is PD for all $i\in \mathbf{S}$, $Q(t)$ will remain PD in any finite interval starting from $t=t_{0}+T_i$, i.e. $Q(t)>0 \hspace{0.1cm}\forall \hspace{0.1cm} t \in [t_{0}+T_i,t_f] \cap \bar{t}_i,$ for any $t_{0}+T_i<t_f<\infty$,
where $t_0+T_i,\hspace{0.1cm}t_f\in \bar{t}_i$ and $\bar{t}_i=\{t \hspace{0.1cm} | \hspace{0.1cm} \sigma(t)=i \}$.}\\
Lemmas 1-2 indicate that the IIE condition on $Z_f(t)$ for any $i\in \mathbf{S}$ can be verified online by checking the determinant of $Q(t)$ online; a positive value implying that the IIE condition on $Z_f(t)$ for active subsystem is satisfied.
\begin{figure}[h]
 \begin{center}
     \includegraphics[width=8.5cm, height=6cm]{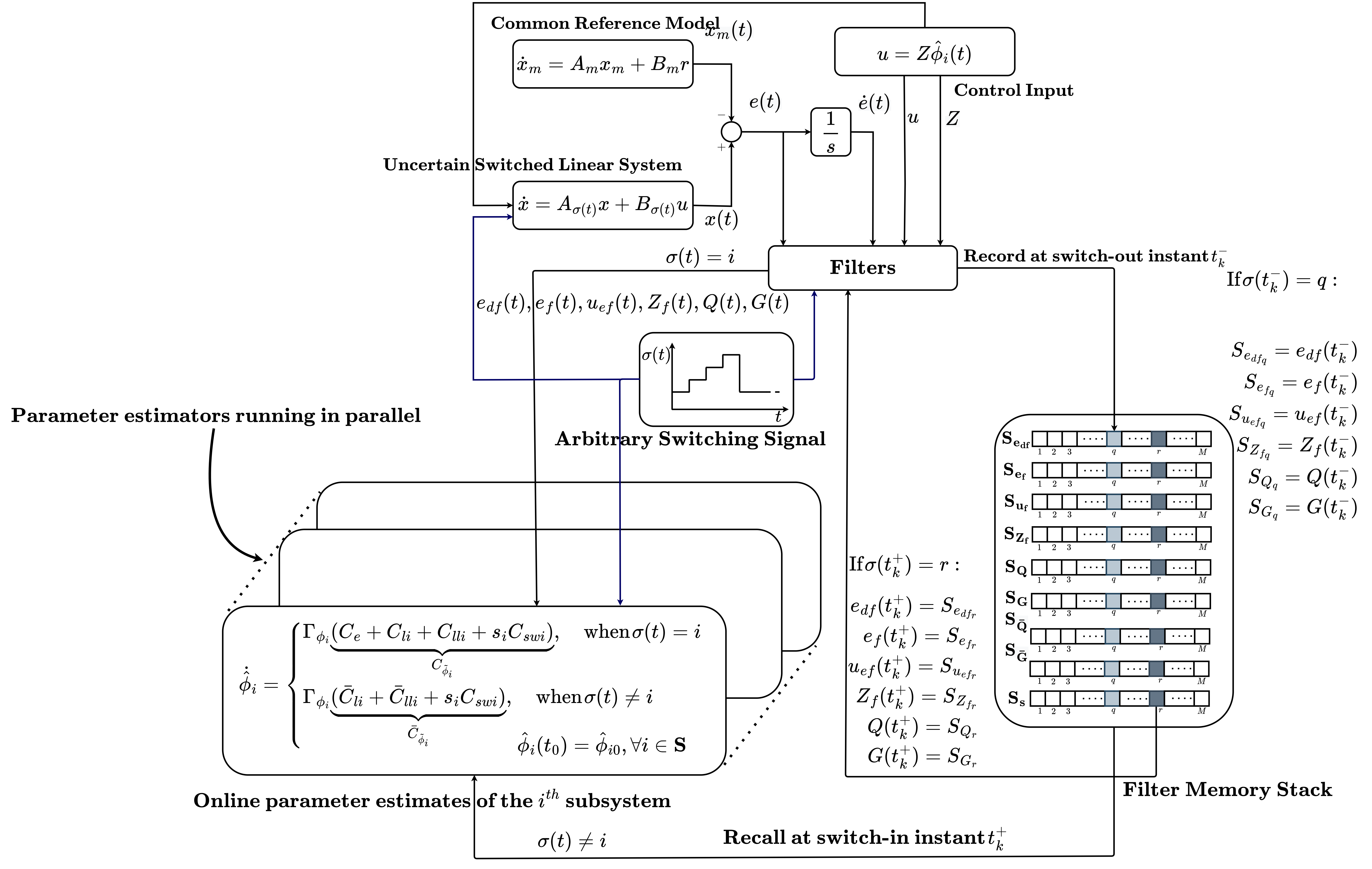}
    \caption{Information flow block diagram for switched MRAC with memory}
    \label{fig:my_label}
\end{center}
\end{figure}
\section{STABILITY ANALYSIS}
\textit{\textbf{Theorem 1}. For the system in (\ref{eq1}), the control law (\ref{control_input}), the parameter estimation law (\ref{eq_para_est}) and any arbitrary switching signal $\sigma(t)$, the origin of the overall error dynamics of the switched system  i.e. $\xi(t) \triangleq [e^T,\tilde{\phi}^T_{1},\tilde{\phi}^T_{2}...,\tilde{\phi}^T_{M}]^T$ is uniformly globally stable. In addition, if Assumption 1 holds, the error ${\xi}(t)$ is uniformly globally exponentially stable (UGES) in the delayed sense for $t\geq t_0+T_{f}$, (where $T_f=\underset{i}{\operatorname{max}}\{T_i\}, \forall i \in \mathbf{S}$) i.e.
\begin{multline}\label{eqc5}
    ||{\xi}(t)||\leq \gamma_1 ||{\xi}(t_0+T_f)||\exp{\{-\gamma_2 (t-t_0-T_f)\}}
\end{multline}
for some positive scalars $\gamma_1, \gamma_2$ independent of initial conditions, provided the following gain condition is satisfied for subsystem $i$.
\begin{equation} \label{eqc2}
    k_{swi}\lambda_{\min}(\mathbf{S}_{\closure{Q}_i})\geq \eta_i
\end{equation}
where $\lambda_{\min}(\bullet)$ denotes the minimum eigenvalue of the argument matrix, and the scalar $\eta_i>0$ is a free parameter, used to alter the rate of convergence.}\\
\textbf{Proof}: Consider the common Lyapunov function candidate
\begin{equation}\label{eq38}
    V=\frac{1}{2}e^TPe+\frac{1}{2}\sum_{i=1}^{M}\tilde{\phi}_{i}^T\Gamma^{-1}_{\phi_{i}}\tilde{\phi}_{i},\hspace{0.6cm} \forall i \in \mathbf{S}
\end{equation}
The Lyapunov function candidate in (\ref{eq38}) satisfies the following inequality.
\begin{equation}\label{eq48}
    \frac{1}{2}\lambda_{m}||{\xi}||^2\leq V\leq\frac{1}{2} \lambda_{M} ||{\xi}||^2
\end{equation}
where the positive constants $\lambda_{m}$, and $\lambda_{M}$ are defined as
\begin{subequations}
\begin{align}
    \lambda_{m}&\triangleq \underset{i}{\operatorname{min}}(\lambda_{\min}(P), \lambda_{\min}(\Gamma^{-1}_{\phi_i}))\\
    \lambda_{M} &\triangleq \underset{i}{\operatorname{max}}(\lambda_{\max}(P),\lambda_{\max}( \Gamma^{-1}_{\phi_i}))
    \end{align}
\end{subequations}

where $\lambda_{\min}(\bullet)$ and $\lambda_{\max}(\bullet)$ denotes minimum and maximum eigenvalue of the argument matrix, respectively. Taking the time derivative of (\ref{eq38}) along the system trajectories yields
\begin{multline}\label{eqc1}
    \dot{V}=-\frac{1}{2}e^TQ_me+\sum_{i=1}^M\mathfrak{I}_i[\tilde{\phi}_{i}^TZ^TB_i^TP^Te+\{\tilde{\phi}_{i}^T (Z e^T P B_i\\+k_{li}Z_f^T(Z_f\phi_{i}-Z_f\hat{\phi}_{i}(t))+k_{lli}(Q\phi_{i}-Q\hat{\phi}_{i}(t))\\+k_{swi}(\mathbf{S}_{\closure{Q}_i}\phi_{i}-\mathbf{S}_{\closure{Q}_i}\hat{\phi}_{i}(t)))\}]+V_{id},\hspace{1cm} \forall i \in \mathbf{S}
\end{multline}
where $Q_m$ and $V_{id}$ are defined as $-Q_m=A_{m}^TP+PA_{m}$ and 
    $V_{id}=-\sum_{\substack{i=1}}^{M}(1-\mathfrak{I}_i)[k_{li}\underbrace{\tilde{\phi}_{i}^T\mathbf{S^T_{Z_{f_i}}}}_{\mathbf{{S_{\varepsilon}}}_{i}(t)}\mathbf{S_{Z_{f_i}}}\tilde{\phi}_{i}+k_{lli}\tilde{\phi}_{i}^T\mathbf{S_{Q_{i}}}\tilde{\phi}_{i}\\+s_ik_{swi}\tilde{\phi}_{i}^T\mathbf{S}_{\closure{Q}_i}\tilde{\phi}_{i}],\hspace{0.2cm} \forall i \in \mathbf{S}$ respectively, where $\mathbf{{S_{\varepsilon_i}}}^T(t) \in \mathbf{R}^m$ is known as the memory prediction error \cite{https://doi.org/10.48550/arxiv.2204.03338}, then
(\ref{eqc1}) can be written as
\begin{equation}\label{eq60c}
    \dot{V}=-\frac{1}{2}e^TQ_me+V_{ad}+V_{id} 
\end{equation}
where $V_{ad}=-\sum_{i=1}^M\mathfrak{I}_i[k_{li}\underbrace{\tilde{\phi}_{i}^TZ_f^T}_{\varepsilon_i(t)}Z_f\tilde{\phi}_{i}+k_{lli}\tilde{\phi}_{i}^TQ\tilde{\phi}_{i}\\+s_ik_{swi}\tilde{\phi}_{i}^T\mathbf{S_{\closure{Q}_i}}\tilde{\phi}_i]$, where $\varepsilon^T_{i}(t)\in \mathbf{R}^{m}$ is typically known as the prediction error \cite{slotine1989composite}. Let $ \lambda_{mZi}=\lambda_{\min}(\mathbf{S^T_{Z_{f_i}}}\mathbf{S_{Z_{f_i}}})$ and $ \lambda_{mQi}=\lambda_{\min}(\mathbf{S_{Q_i}})$ then (\ref{eq60c}) can be written as
\begin{multline}\label{eq60}
    \dot{V}\leq -\frac{1}{2}e^TQ_me-\sum_{i=1}^M\mathfrak{I}_i[k_{li}||\varepsilon_i(t)||^2+k_{lli}\tilde{\phi}_{i}^TQ\tilde{\phi}_{i}\\+s_i\eta_i||\tilde{\phi}_{i}||^2]-\sum_{\substack{i=1}}^{M}(1-\mathfrak{I}_i)(\lambda_{mZi}||\tilde{\phi}_{i}||^2+\lambda_{mQi}||\tilde{\phi}_{i}||^2\\+s_i\eta_i ||\tilde{\phi}_{i}||^2) \hspace{2cm} \forall i\in \mathbf{S}
\end{multline}
Depending on whether the IIE condition holds for the active subsystem, two cases are possible\\\textit{\textbf{Case 1}}: When $t<t_0+T_f$ (i.e. some subsystems fulfills the IIE condition) :\\
Using property 1, $\dot{V}$ can be upper bounded as
\begin{equation}\label{eq42}
    \dot{V}\leq -\frac{1}{2}\lambda_{min}(Q_m)||e(t)||^2-\sum_{i=1}^M\mathfrak{I}_ik_{li}||\varepsilon_i(t)||^2 
\end{equation}
\hspace{6cm}$\forall t < t_0+T_f$

From (\ref{eq42}), $\dot{V}$ is negative semi-definite and hence, the error ${\xi}(t)$ is uniformly globally stable (UGS) $\forall t <t_0+T_f$ \cite{khalil2002nonlinear}.\\ \textit{\textbf{Case 2}}: When $t\geq t_0+T_f$ (i.e. all subsystems satisfy the IIE condition) :\\
This case indicates that Assumption 1 holds for all subsystems, therefore, for $t\geq t_{0}+T_f$, $\dot{V}$ in (\ref{eq60}) can be upper bounded using Lemma 2, and (\ref{eq_Q_bar}) along with the gain condition (\ref{eqc2}) as
\begin{subequations}
\begin{align}
    \dot{V} &  \leq -\frac{1}{2}\lambda_{\min}(Q_m)||e(t)||^2 -\sum_{i=1}^M\mathfrak{I}_i[k_{li}||\varepsilon_i(t)||^2 \nonumber\\
    &-k_{lli}\text{exp}\{-k_s (t-t_0)\}\gamma_i||\tilde{\phi}_{i}||^2-\eta_i||\tilde{\phi}_{i}||^2] \nonumber
    \\
    &-\sum_{\substack{i=1}}^{M}(1-\mathfrak{I}_i)(\lambda_{mZi}||\tilde{\phi}_{i}||^2+\lambda_{mQi}||\tilde{\phi}_{i}||^2\\&\hspace{2cm}+\eta_i ||\tilde{\phi}_{i}||^2), \hspace{0.2cm}\forall t \\
    \dot{V} &  \leq -\frac{1}{2}\lambda_{\min}(Q_m)||e(t)||^2 -\sum_{i=1}^M\mathfrak{I}_i\eta_i||\tilde{\phi}_{i}||^2 \nonumber
    \\
    &-\sum_{\substack{i=1}}^{M}(1-\mathfrak{I}_i)\min(\lambda_{mZi}, \lambda_{mQi}, \eta_i)||\tilde{\phi}_{i}||^2, \hspace{0.2cm}\forall t \geq t_0+T_f \\
    \dot{V} &\leq -\alpha V, \hspace{0.2cm} \forall t\geq t_0+T_f \label{eq57c}
    \end{align}
\end{subequations}
where $\alpha =\frac{\min(\lambda_{\min}(Q_m),\varrho_a,\varrho_i)}{\lambda_{M}}$ with $\varrho_a=2\sum_{i=1}^M\mathfrak{I}_i\eta_i$ and\\
    $\varrho_i=2 \sum_{\substack{i=1}}^{M}(1-\mathfrak{I}_i)\min(\lambda_{mZi},\lambda_{mQi},\eta_i)$\\
Using the Comparison Lemma (Lemma 3.4 of \cite{khalil2002nonlinear}), the differential inequality in (\ref{eq57c}) leads to the subsequent exponentially convergent bound.
\begin{equation}\label{eqc9}
    V(t)\leq V(t_0+T_f)\exp\{-\alpha(t-t_0-T_f)\}, \hspace{0.3cm} \forall t \geq t_0+T_f
\end{equation}
Using (\ref{eq48}), the inequality in (\ref{eqc9}) can be converted to
\begin{multline}\label{eqc10}
    ||
    \xi(t)||\leq \gamma_1||\xi(t_0+T_f)||\exp\{-\gamma_2(t-t_0-T_f)\}\\ \hspace{4cm} \forall t \geq t_0+T_f
\end{multline}
Comparing (\ref{eqc5}) and (\ref{eqc10}), $\gamma_1=\sqrt{\frac{\lambda_{M}}{\lambda_{m}}},$ and $\gamma_2=\frac{\alpha}{2}$. Since, the Lyapunov function in (\ref{eq38}) is radially unbounded and the constants $\gamma_1,\gamma_2$ are independent of initial conditions, the algebraic inequality in (\ref{eqc10}) proves UGES (in a delayed sense) of the error $\xi(t), \forall t\geq t_0+T_f$.\\

\section{SIMULATION RESULT}
Consider the following subsystem matrices for the switched system
\begin{subequations}\label{eq66}
\begin{align}
A_1&=\begin{bmatrix}
0 & 1;
-5 & -6
\end{bmatrix},
&B_1=\begin{bmatrix}
0;
1 
\end{bmatrix}\\
A_2&=\begin{bmatrix}
0 & 1;
-5.5 & -6.5
\end{bmatrix},
&B_2=\begin{bmatrix}
0;
1 
\end{bmatrix}\\
A_3&=\begin{bmatrix}
0 & 1;
-6 & -7
\end{bmatrix},
&B_3=\begin{bmatrix}
0;
1 
\end{bmatrix}\\
A_4&=\begin{bmatrix}
0 & 1;
-8 & -9
\end{bmatrix},
&B_4=\begin{bmatrix}
0;
1 
\end{bmatrix}
\end{align}
\end{subequations}
and common reference system dynamics is given by $A_m=\begin{bmatrix}
    0 & 1;
-3 & -4
    \end{bmatrix},
    B_m=\begin{bmatrix}
0;
1 
\end{bmatrix}$
and a switching signal with a switching interval $\{t_{k+1}-t_k\}_{\forall k \in \mathbf{N}}=30$ sec is chosen and reference signal is given by $r(t)=\bar{r}(t)+\delta_s(t)$, where $\bar{r}(t)$ is the desired reference signal and $\delta_s(t)$ is used to satisfy the IIE condition at every switching instant, given by $\bar{r}(t)\footnote{Constant value is taken to represent a non PE signal.}=0$ and $\delta_s(t-t_s)\footnote{$\delta_s(t-t_s)$ is exponentially decaying signal hence it is a non PE signal.}=10e^{-0.1(t-t_s)}(\sin(2(t-t_s))+\sin(3(t-t_s))+\sin(4(t-t_s))+\sin(5(t-t_s))+\sin(6(t-t_s))$
where $t_s=t_k \hspace{0.2cm} \text{when} \hspace{0.2cm} t_k\leq t <t_{k+1}, \hspace {0.2cm}\forall k \in \mathbf{N}$. The design parameters are chosen as $\Gamma_{\theta i}=I_2, \hspace{0.1cm} k_{li}=k_{lli}=k_{swi}=k_f=k_s=1, \forall i \in \mathbf{S}$.

\bibliographystyle{ieeetr}
\footnotesize{\bibliography{LCSS}}
\end{document}